\documentclass[twocolumn,showpacs,preprintnumbers,amsmath,amssymb,aps, prb,floatfix,groupedaddress,superscriptaddress]{revtex4}
\usepackage{graphicx} 
\usepackage{subfigure,tabularx}
\usepackage{dcolumn} 
\usepackage{bm} 
\usepackage{epsfig}
\usepackage{color}
\usepackage{multirow,longtable}

\begin{document}

\title{PuPt$_2$In$_7$: a computational and experimental investigation}

\author{H. B. Rhee}
\affiliation{Department of Physics, University of California, Davis, CA 95616, USA}

\author{F. Ronning}
\affiliation{Los Alamos National Laboratory, Los Alamos, New Mexico 87545, USA}

\author{J.-X. Zhu}
\affiliation{Los Alamos National Laboratory, Los Alamos, New Mexico 87545, USA}

\author{E. D. Bauer}
\affiliation{Los Alamos National Laboratory, Los Alamos, New Mexico 87545, USA}

\author{J. N. Mitchell}
\affiliation{Los Alamos National Laboratory, Los Alamos, New Mexico 87545, USA}

\author{P. H. Tobash}
\affiliation{Los Alamos National Laboratory, Los Alamos, New Mexico 87545, USA}

\author{B. L. Scott}
\affiliation{Los Alamos National Laboratory, Los Alamos, New Mexico 87545, USA}

\author{J. D. Thompson}
\affiliation{Los Alamos National Laboratory, Los Alamos, New Mexico 87545, USA}

\author{Yu Jiang}
\affiliation{Chemical Sciences Division,
Lawrence Berkeley National Laboratory,
Berkeley, California 94720, USA}
\author{C. H. Booth}
\affiliation{Chemical Sciences Division,
Lawrence Berkeley National Laboratory,
Berkeley, California 94720, USA}

\author{W. E. Pickett}
\affiliation{Department of Physics, University of California, Davis, CA 95616, USA}

\date{\today}

\begin{abstract}
Flux-grown single crystals of PuPt$_2$In$_7$ are characterized and found to be both non-superconducting and non-magnetic down to 2 K. The Sommerfeld specific heat coefficient of $~ 250$ mJ/mol K$^2$ indicates heavy fermion behavior. We report the results of generalized gradient approximation (GGA)+$U$ calculations of PuPt$_2$In$_7$ and as yet unsynthesized isovalent PuPt$_2$Ga$_7$. The strength of the $c$-$f$ hybridization of PuPt$_2$In$_7$ is similar to the PuCoIn$_5$ superconductor. The bare and $f$-weighted susceptibility within the constant-matrix-element approximation is calculated, showing a maximum along the $q_z$ direction at $q_x = q_y = 0.5$. A similar and slightly stronger maximum is also found in the structurally related heavy-fermion materials PuCoGa$_5$ and PuCoIn$_5$. The absence of superconductivity in PuPt$_2$In$_7$ is examined based on the results of our calculations.
\end{abstract}

\maketitle

\section{Introduction}
\begin{figure}[bt]
\begin{center}
\includegraphics[draft=false,width=\columnwidth]{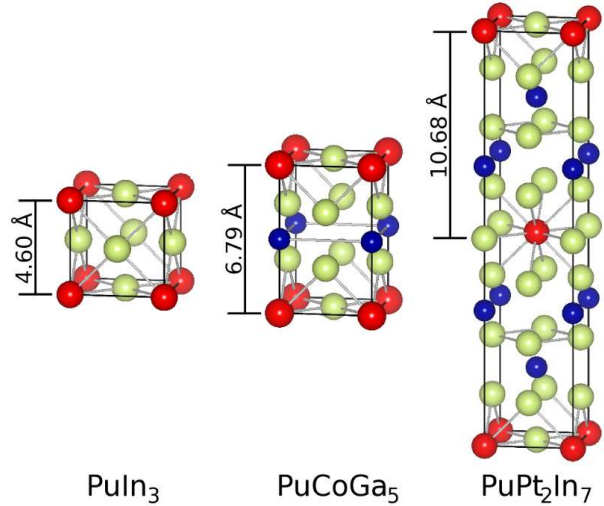}
\end{center}
\caption{(Color online.) Crystal structures, obtained using {\sc VESTA},\cite{vesta} of some Pu-based compounds. The Pu atoms are colored red, Pt/Co atoms dark blue, and In/Ga atoms light green.}
\label{structure}
\end{figure}

Magnetically mediated superconductivity in heavy-electron systems, specifically Ce and U compounds, has been known to exist for over 30 years.\cite{steglich, ott, stewart, jaccard} Among the known heavy fermion superconductors a particularly rich family includes the so-called ``115,'' ``127,'' and ``218'' structures which are all variants of the ``103'' parent compound, crystallizing in the Ho$_m$Co$_n$Ga$_{3m+2n}$ architecture (see Fig.~\ref{structure}).\cite{Thompson} Many of these compounds are known to be superconducting.\cite{mathur, hegger, petrovic, petrovic2, chen, sarrao, wastin, kac, bauer, bauer3} It is widely expected that in these systems, spin fluctuations are what bind the Cooper pairs, and the balance between these local-moment fluctuations and long-range magnetism in the vicinity of a quantum critical point is crucial for superconductivity to take place. There are however compounds that belong in this structural family but do not superconduct, as they tend to shy away from this ideal balance. For example, the $f$~electrons in the U-115s, -218s, and Np-115s are too itinerant to exhibit superconductivity,\cite{sechovsky,kaneko,metoki,elgazzar,aoki,aoki2} and in AmCoGa$_5$ they are too localized.\cite{opahle,javorsky} CeRhIn$_5$, CePt$_2$In$_7$, and Ce$_2$RhIn$_8$, all nonsuperconducting antiferromagnets at ambient pressure, require compression to delocalize the $f$~electrons and make them available for electron-electron pairing.\cite{hegger,bauer,nicklas} Pu-based compounds are particularly interesting, because within the actinides it is Pu that straddles the line between bearing localized and itinerant $5f$ electron states.

Pu compounds are often considered the hole analog of their Ce counterparts because they have five $f$ electrons in the $5f_{5/2}$ spin-orbit split multiplet.  In fact, the two 115 subgroups manifest very similar behaviors from their Curie--Weiss-like magnetic susceptibilities\cite{wastin,hegger,curro,pagliuso} to their quasi-two-dimensional Fermi surfaces (FSs).\cite{oppeneer, hall, settai} The In-bearing members show remarkably similar properties as well: PuIn$_3$ is a 14~K antiferromagnet,\cite{PuIn3} while CeIn$_3$ is a 10~K antiferromagnet,\cite{mathur} and PuCoIn$_5$ and CeCoIn$_5$ are both 2~K superconductors.\cite{bauer3, petrovic} Why the $T_c$'s of the PuCoGa$_5$ and PuRhGa$_5$ superconductors are so much higher however remains elusive.

Based on the impressive list of superconductors discovered in the past 25 years with two-dimensional (2D) structures and properties, a guideline can be made that some aspects of 2D structures make them more favorable for superconductivity and will therefore give rise to a higher $T_c$. The average spin fluctuation frequency is higher in quasi-2D systems than 3D, and this brings about a larger Cooper pairing energy.\cite{monthoux99} Indeed, the PuRh$_{1-x}$Co$_x$Ga$_5$ ($0 \le x \le 1$) compounds with $T_c$ values up to 18 K follow a linear relation in $T_c$ vs.\ the axial ratio $c/a$,\cite{bauer2} which is also observed in Ce$M$In$_5$ (and, interestingly, with an almost identical slope to that of Pu-based cousins).\cite{pagliuso,kumar} Recently, CePt$_2$In$_7$---a structurally and electronically more 2D version of 115---was discovered.\cite{bauer, altarawneh} Although $T_c$ was not enhanced, it did achieve a maximum superconducting transition temperature of 2.1~K, comparable to the other Ce-based 115s'.

In this paper, we report the discovery of the Pu analog to CePt$_2$In$_7$. We find that PuPt$_2$In$_7$ is a mass enhanced paramagnet which lacks superconductivity down to 2~K. We report electronic structure calculations on PuPt$_2$In$_7$, including densities of states, band structures, and Fermi surfaces. We present also analogous analyses on isovalent PuPt$_2$Ga$_7$, which has yet to be synthesized. In addition, we have calculated the constant-matrix-element and atomic-character-matrix-element noninteracting magnetic susceptibilities of PuPt$_2$In$_7$ and PuPt$_2$Ga$_7$, and of PuCoGa$_5$ and PuCoIn$_5$ as points of comparison. While the Fermi surfaces of the 127 compounds are qualitatively distinct from the 115s, all four Pu compounds exhibit a row of peaks in the susceptibility along the $q_z$ direction at $q_x = q_y = 0.5$. We discuss the possible implication of these results for understanding Pu-based superconductivity.

\begin{table}[t]
\caption{Table of structural parameters and atomic positions for PuPt$_2$In$_7$ determined from single crystal x-ray diffraction.}
\begin{ruledtabular}
\begin{tabular}{lccccccc}
Space group & \textit{I}4/\textit{mmm} \\
$a$ (\AA) & 4.5575(7) \\
$b$ (\AA) & 4.5575(7) \\
$c$ (\AA) & 21.362(6) \\
Volume (\AA$^3$) & 443.71(16) \\
Formula units/cell & $Z=2$ \\
\\
Atom & $x$ & $y$ & $z$
\\\hline
Pu & 0 & 0 & 0 \\
Pt & 0 & 0 & 0.32626(6) \\
In1 & 0 & 0 & 0.5 \\
In2 & 0 & 0.5 & 0.2500 \\
In3 & 0 & 0.5 & 0.10597(11) \\
\end{tabular}
\label{tbl:latts}
\end{ruledtabular}
\end{table}

\section{Experiment}
Single crystals of PuPt$_2$In$_7$ were grown by the self flux method from the respective elements with an excess of In metal.  The reactions were loaded in the ratio Pu:Pt:In (1:4:30) using 2 cm$^3$ alumina crucibles which were sealed under vacuum in quartz ampoules.  The isolated single crystals grew with a plate-like habit and were found to be PuPt$_2$In$_7$ based on single crystal x-ray diffraction analysis. The single crystal x-ray data were collected on a Bruker D8 equipped with a APEX2 CCD detector.  Full spheres of data were collected at room temperature and the collections were handled in batch runs at different $\omega$  and $\phi$  angles.  The structure was refined using the atomic coordinates from the isostructural CePt$_2$In$_7$ compound.  The data integration and refinement procedures were completed using SAINT-Plus, SHELXS97, and SHELXL97 programs. PuPt$_2$In$_7$ stabilizes into a body-centered tetragonal structure (see Fig.~\ref{structure} and Table~\ref{tbl:latts}). While in PuCoGa$_5$ the PuGa$_3$ layer and the CoGa$_2$ layer stack alternately, resulting in a primitive structure, PuPt$_2$In$_7$ has two layers of PtIn$_2$ for each PuIn$_3$.

\begin{figure}[bth]
\vbox{
\includegraphics[width=2.8in,angle=0]{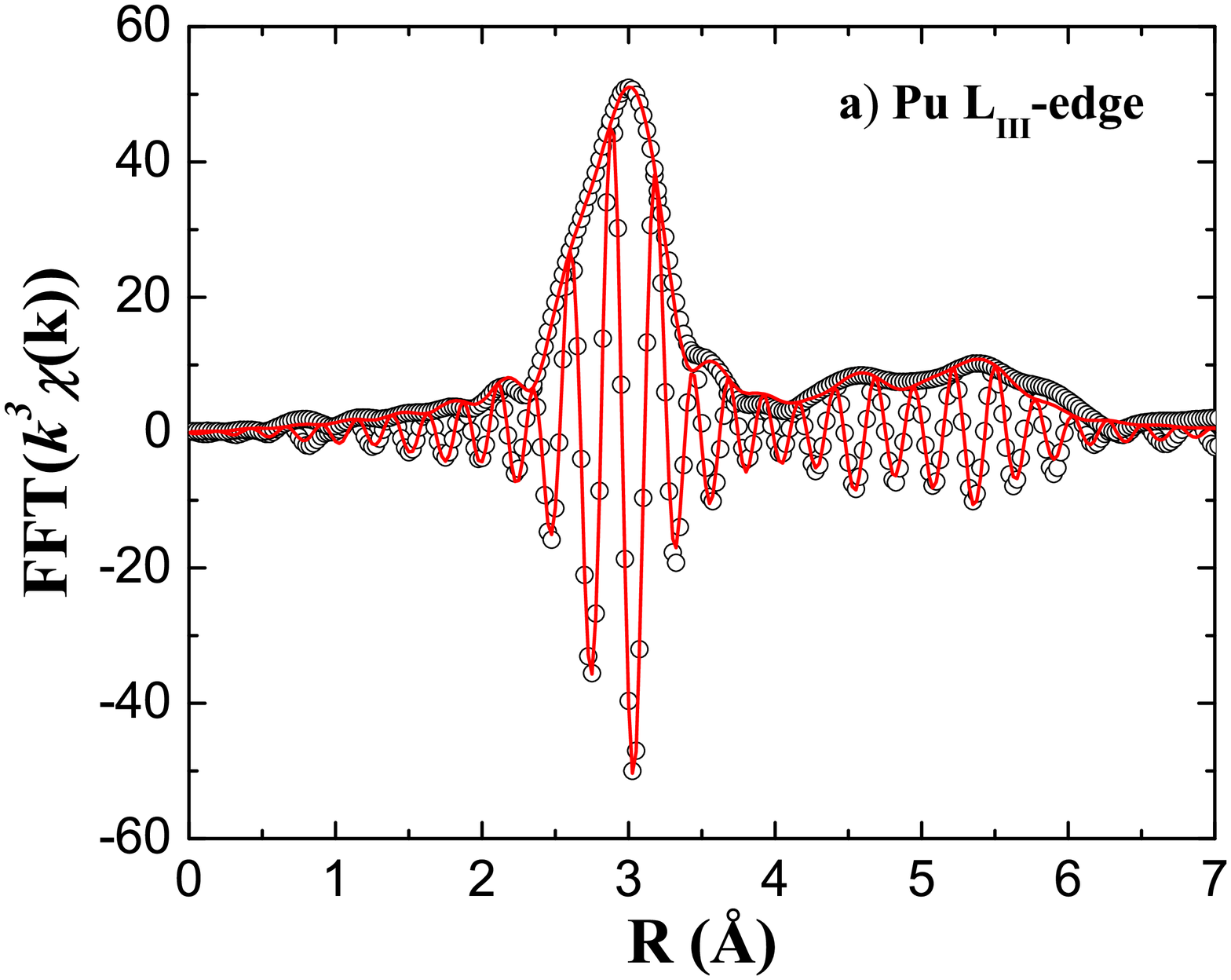}
\includegraphics[width=2.8in,angle=0]{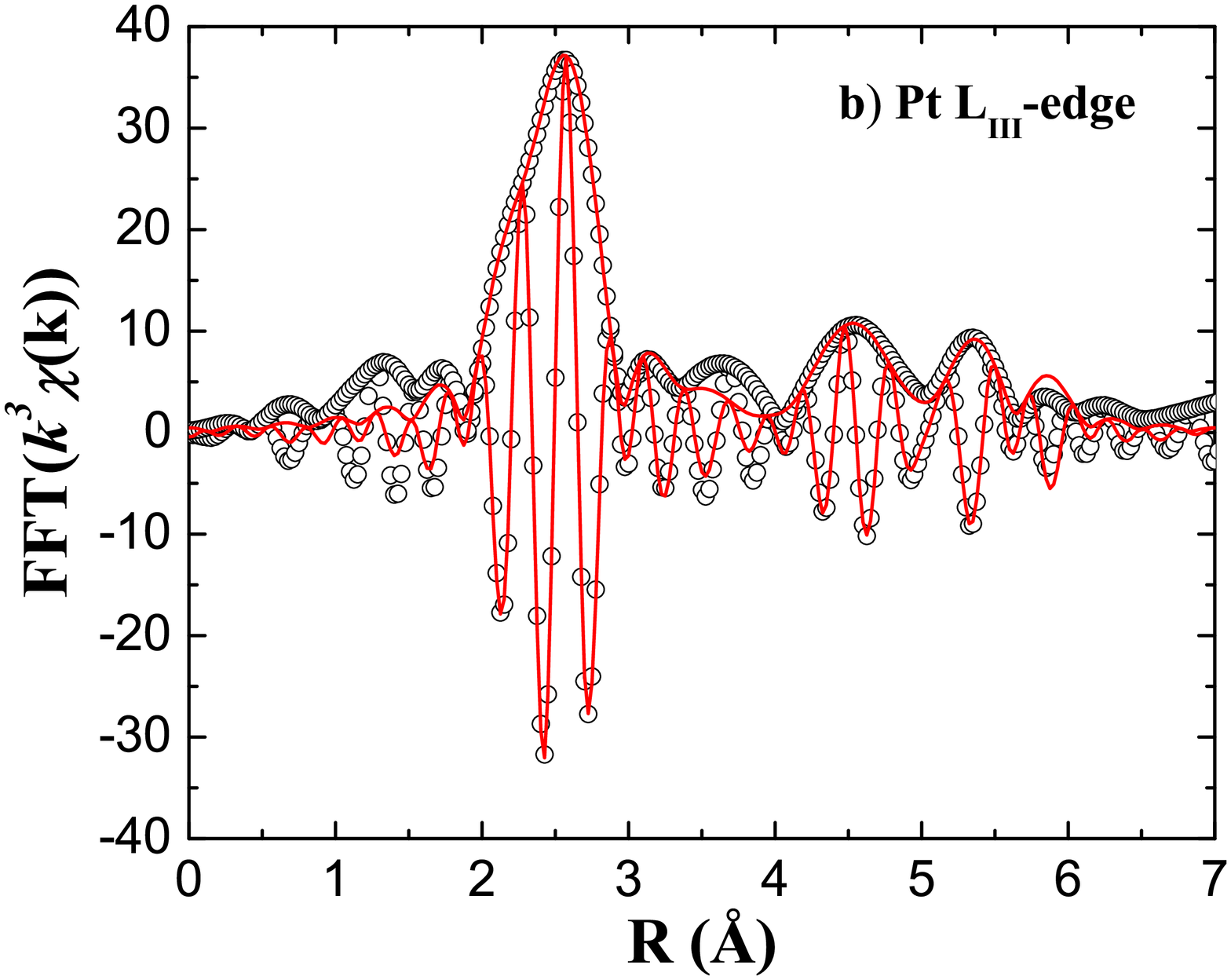}
}
\caption{Fast Fourier-transformed $r$-space data (open symbol) and the fit
(solid line) for (a) Pu $L_{\rm III}$-edge and (b) Pt $L_{\rm III}$-edge. Both
EXAFS data sets were measured at $T = 30$ K, with the Fourier-transformed $k$ range of 3.5--13.5
\AA$^{-1}$ and a Gaussian window of 0.3 \AA$^{-1}$. The $r$-space fit range
is 2.1--5.4 \AA{} for both edges. Here only the real part, Re, and the
amplitude, $\sqrt{\text{Re}^2+\text{Im}^2}$, of FFT($k^3\chi(k)$) are plotted.}
\label{fig-rs}
\end{figure}

\begin{table*}[bth]
\begin{minipage}{0.85\textwidth}
\caption{EXAFS fit results for the Pu and Pt $L_{\rm III}$-edges on PuPt$_2$In$_7$.  Fit- and Fourier-transform ranges are listed in the caption of Fig.~\ref{fig-rs}.  Though we only show single-scattering peaks shorter than 5.0 \AA, all single- and multiple-scattering peaks within the fit range are included. To obtain a better estimate of the contribution from the farther atoms in the fit range, the single-scattering peaks between 5.4 and 6.0 \AA\ are also included in the fit and are held together with one single $\sigma^2$. Coordination numbers $N$ are held fixed to the nominal structure. A small vibration of the lattice is allowed by constraining the shifts of all longer bonds to the shortest bonds and keeping the shortest ones free to move.  In addition, in the Pt edge fit, the Pt-Pu pairs at $\sim$4.92 \AA{} are fixed to
the Pu-Pt pairs with the same $R$, and $\sigma^2$ to reduce the fitting
parameter. $S_0^2$, $\Delta E$, and the fit quality are 0.90(1)~eV, -10.0(1)~eV, and 7.6\% for the Pu edge, respectively, and 0.90(1)~eV, -8.1(15)~eV, and 18.20\% for the Pt edge. [Note that the bad fit quality for the Pt edge fit and large uncertainty in the correlated Debye fit for the Pt-In(1) pair are caused by the oscillation around 3.5~\AA.] The number of free parameters in the fits is 14 for the Pu and 15 for the Pt edge, far below the number of independent data points as given by Stern's rule,\cite{Stern93} which is $\sim$23 for both fits.}
\begin{ruledtabular}
\begin{tabular}{lccccc}
&     & $\sigma^{2}$ & $R$ & $\sigma^{2}_\mathrm{stat}$ & $\theta_\mathrm{cD}$ \\
& $N$ & (\AA$^2$) & (\AA) & (\AA$^2$) & (T) \\
\colrule
 Pu-In(3)/In(1)
& 12 & 0.002( 3) & 3.224(4) &  0.0001(1) &  211(2) \\
 Pu-Pu
& 4 & 0.004(2) & 4.567 &  0.0015(5) & 145(9) \\
 Pu-Pt
& 8 & 0.005(3) & 4.925 &  0.0039(4) & 238(15) \\
 Pt-In(3)/In(2)
& 8 & 0.0003(2) & 2.745(6) &  -0.0003(5) & 266(29) \\
 Pt-In(1)
& 1 & 0.0003 & 3.702 &  0.0009(18) & 408(368) \\
 Pt-Pt
& 8 & 0.0006(5) & 4.561 & -0.0001(8) & 255(33) \\
\colrule
site-interchange & Pu/In(1) & Pu/In(2) & Pu/In(3) & Pu/Pt & Pt/In(1) \\
 fraction (\%) & 6$\pm$4 & 15$\pm$6 & 0$\pm$4 & 3$\pm$5 & 18$\pm$21  \\
\end{tabular}
\end{ruledtabular}
\end{minipage}
\label{tab-debye}
\end{table*}

To understand the local structure of PuPt$_2$In$_7$, fluorescence extended x-ray
absorption fine structure (EXAFS) data were collected at the Stanford
Synchrotron Radiation Lightsource (SSRL) on the Pu and Pt $L_{\rm III}$-edges at
beamline 11-2, using a half-tuned double crystal Si(220) monochromator, with a
slit height of 0.6 and 0.5~mm for the measurement of each edge,
respectively. A six-month old crystal sample was triply contained in a sample
holder with Kapton windows, and was placed 45$^{\circ}$ to the incident x-ray beam. The self-absorption corrected EXAFS data are reduced using standard procedures outlined in Refs.~\onlinecite{hayes} and \onlinecite{li}, including fitting an embedded-atom absorption function $\mu_0(E)$ using a seven-knot cubic spline function with a maximum photoelectron wave vector $k$ of 15 \AA$^{-1}$. The EXAFS function is then defined as $\mu(k)/\mu_0(k)-1$, where $\mu$ is the absorption coefficient, $k=\sqrt{(2m_e/\hbar^2)(E-E_0)}$ is the photoelectron wave vector, $m_e$ is the electron rest mass, $E$ is the incident energy, and $E_0$ is the absorption edge threshold energy, which is defined arbitrarily to be the half height of the edge and allowed to vary in the fit.

$k^3$-weighted EXAFS data, $k^3\chi(k)$, are fast Fourier transformed (FFT) to
$r$ space (FFT($k^3\chi(k)$)), with a FFT range of $k = 3.5$--13.5
\AA$^{-1}$ and a Gaussian window of 0.3 \AA$^{-1}$, for both Pu and Pt edges. The
$r$-space EXAFS data are then fit with theoretical {\it FEFF} functions\cite{ankudinov}
calculated based on the $I4/mmm$ lattice structure. The $r$-space data versus fit
are shown in Fig.~\ref{fig-rs}; the Debye-Waller factors, $\sigma^2(T)$, for
some atom pairs ($< 5$ \AA) are fit to the correlated Debye
model\cite{crozier} to obtain the static distortion, $\sigma^2_\mathrm{stat}$, and
the correlated Debye temperature, $\theta_\mathrm{cD}$ (shown in Table
\ref{tab-debye}).  The Pu occupancy ($\sim$98$\pm$16\%) is estimated by
allowing the amplitude of the Pu-Pu peak (4.56 \AA) to vary in the Pu edge fit,
though the fit quality does not change from the previous fit, which assumes 100\%
Pu occupancy. By arbitrarily constraining $\sigma^2_\mathrm{stat} \ge 0$ for the Pu-Pu
pair, the Pu occupancy is estimated to be $>$~83\%.  Possible ion/ion site
interchange, such as Pu to In(1,2,3), and Pt to In(1), are also examined
using a similar method to that in Ref.~\onlinecite{Bauer02}. From these fits, the
percentage of Pu site-interchange with other ions, shown in the lower part of Table
\ref{tab-debye}, is estimated to be close to zero within a small error.  Hence,
the fit results indicate well ordered local lattice structure around both Pu
and Pt ions.

\begin{figure}[tbh]
\begin{center}
\includegraphics[draft=false,width=\columnwidth]{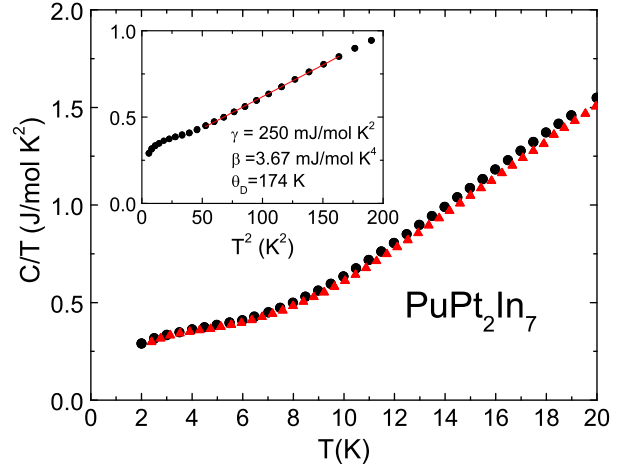}
\end{center}
\caption{(Color online) Specific heat data ($C/T$) vs.\ temperature of PuPt$_2$In$_7$, taken in zero field (black circles) and 6 T (red triangles). Inset shows the data plotted vs.\ $T^2$ along with a linear fit between 7 and 13 K, from which estimates of the Sommerfeld coefficient and Debye temperature were obtained.}
\label{cp}
\end{figure}

Specific heat data are shown in Fig.~\ref{cp}. A fit of the data to $C/T = \gamma + \beta T^2$ between 7 and 13 K gives an enhanced Sommerfeld coefficient of 250 mJ/mol K$^2$ and $\beta = 3.67$ mJ/mol K$^4$. Using the formula $\Theta_D = (12/5 * \pi^4 n k_B)^{1/3} \beta$---where $k_B$ is the Boltzmann constant and $n$, the number of atoms per formula unit, is equal to 10---we get a Debye temperature, $\Theta_D = 174$ K. The Sommerfeld coefficient is larger than that of PuCoGa$_5$ ($\gamma \simeq 100$ mJ/mol K$^2$). Thus, the value of $\gamma$ for PuPt$_2$In$_7$ likely represents a reduction in the characteristic spin fluctuation temperature of PuPt$_2$In$_7$ relative to PuCoGa$_5$. At temperatures below 7 K, a small hump is seen in the specific heat which may represent short range correlations. Susceptibility measurements down to 2 K (not shown) show no evidence of superconductivity or long-range magnetic order.

\section{Computational Results}
Electronic structure calculations using the generalized gradient approximation (GGA) within density functional theory\cite{dft} were carried out with {\sc WIEN2k},\cite{wien} which employs full-potential linearized augmented planewaves and local orbitals. In principle, GGA is more appropriate than LDA, as exemplified by the GGA studies by Robert, Pasturel, and Siberchicot of Pu compounds.\cite{robert} We adopted the Perdew-Burke-Ernzerhof\cite{pbe} exchange-correlation potential based on the generalized gradient approximation, and we included spin-orbit (SO) interactions through a second variational method. We performed calculations with and without the Hubbard $U$ (using the around mean field double-counting correction\cite{amf}) and exchange $J$; we used the widely accepted values of $U=3$--4 eV and $J=0.6$ eV for Pu.\cite {elgazzar,savrasov,shick,pourovskii,suzuki}

The experimental lattice parameters of PuPt$_2$In$_7$ (see Table~\ref{tbl:latts}) were used. They were also used to estimate the size of the hypothetical compound PuPt$_2$Ga$_7$, by means of extrapolating the lattice differences of PuCoGa$_5$ (Ref.~\onlinecite{sarrao}) and PuCoIn$_5$ (Ref.~\onlinecite{bauer3}). The inferred lattice parameters for PuPt$_2$Ga$_7$ are thus $a = 4.22$~\AA\ and $c = 19.51$~\AA. The same internal parameters for PuPt$_2$In$_7$ were used for PuPt$_2$Ga$_7$.

Paramagnetic (PM), ferromagnetic (FM), and two different antiferromagnetic (AFM) calculations were performed for both PuPt$_2$In$_7$ and PuPt$_2$Ga$_7$, and the relative energies are listed in Table~\ref{tbl:energies}. AFM I represents a configuration in which the antiferromagnetic $\mathbf{q}$-vector is $(1/2,1/2,0)$, and AFM II has a wavevector of $(0,0,1)$. Regardless of the value of $U$, the energy of PM PuPt$_2$In$_7$ stays far above those of the other magnetic configurations, in contrast with experimental observations (although, the difference shrinks with increasing U). Even though the AMF double-counting method was implemented specifically for its suppression of magnetism,\cite{ylvisaker} and has correctly predicted the nonmagnetic ground state for $\delta$-Pu, PuCoGa$_5$, and the Pu-218s \cite{shick-pu,elgazzar,oppeneer} when no other double-counting approach has been successful, it fails to have the same effect on PuPt$_2$In$_7$. A reason for this may be that the distance between the Pu atom and its nearest neighbor is greater in PuPt$_2$In$_7$ (3.2 \AA) than the other compounds (it is 3.0 \AA\ for PuCoGa$_5$, 2.5 \AA\ for Pu$_2$CoGa$_8$, and 2.6 \AA\ for Pu$_2$RhGa$_8$; $\delta$-Pu does not have a ligand but the Pu-Pu distance is 3.1\AA), which would provide more room for larger local moments. In the GGA scheme, the ground-state configuration of PuPt$_2$In$_7$ is AFM I, but the FM and AFM II systems become more stable when $U$ is set to 3 eV. At 4 eV, the AFM II configuration has the lowest energy, with the FM state just 2 meV higher. This indicates that increasing U favors FM interactions within planes and weak AFM interactions between planes.

\begin{table}[bthp]
\caption{Relative total energies (in eV) from GGA and GGA+$U$ calculations of different magnetic configurations of the Pu-127s. The AFM I configuration has a $\mathbf{q}$-vector of $(1/2,1/2,0)$, AFM II has one of $(0,0,1)$. $J=0.6$ eV for all $U \neq 0$ calculations.}
\begin{ruledtabular}
\begin{tabular}{ll|rrrr} 
         && PM    & FM    & AFM I & AFM II\\
\hline
\multirow{3}{*}{PuPt$_2$In$_7$}
&$U=0$ eV & +1.15 & +0.09 & ~0.00 & +0.09 \\
&$U=3$ eV & +0.57 & ~0.00 & +0.01 & ~0.00\\
&$U=4$ eV & +0.17 & +0.002 & +0.04 & ~0.00 \\
\hline
\multirow{3}{*}{PuPt$_2$Ga$_7$}
&$U=0$ eV & +0.93 & +0.10 & ~0.00 & +0.11\\
&$U=3$ eV & +0.41 & +0.03 & ~0.00 & +0.001 \\
&$U=4$ eV & +0.06 & +0.06 & +0.01 & ~0.00 \\
\end{tabular}
\label{tbl:energies}
\end{ruledtabular}
\end{table}

The energies of PuPt$_2$Ga$_7$ at $U=0$ are not unlike those of PuPt$_2$In$_7$, but when $U$ is turned on, competition for the ground state is not between FM and AFM II but the two antiferromagnetic flavors. The general similarities suggest it is likely that PuPt$_2$Ga$_7$ will also be a paramagnet, but with some differences in the strength and character of short-range magnetic correlations and with weak AFM interactions between planes. Similar to calculations of $\delta$-Pu (Ref.~\onlinecite{soderlind}), we find a sizable cancellation of spin and orbital moments. For instance, for the AFM II state with $U = 4$ eV we have a spin moment of 4.262~$\mu_\mathrm B$ and an orbital moment of $-3.404$~$\mu_\mathrm B$.

\section{Electronic structure}
\begin{figure}[tbh]
\begin{center}
\includegraphics[draft=false,width=\columnwidth,clip]{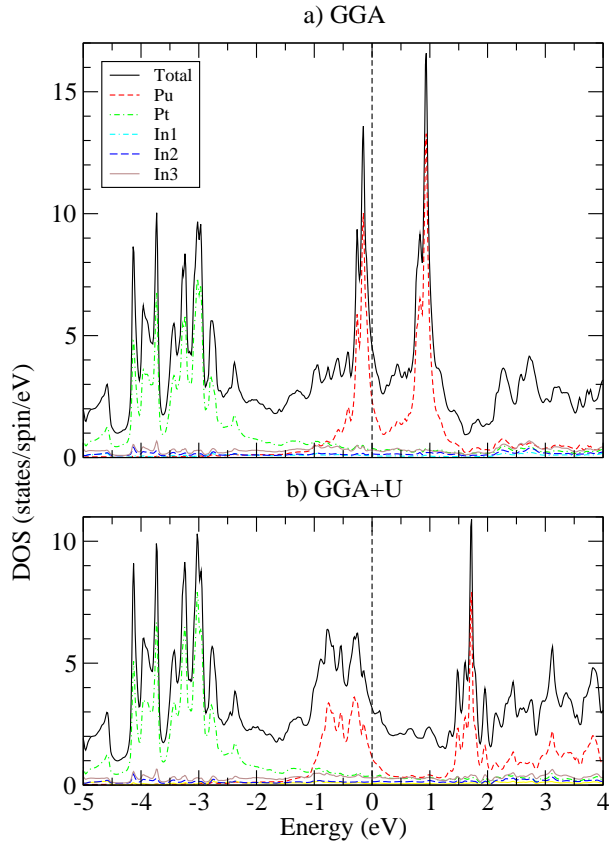}
\end{center}
\caption{(Color online) Total and partial DOSs of PM PuPt$_2$In$_7$ from a) GGA and b) GGA+$U$ ($U=3$ eV, $J=0.6$ eV) calculations.}
\label{in-dos}
\end{figure}

Fig.~\ref{in-dos} shows the calculated density of states (DOS) of paramagnetic PuPt$_2$In$_7$ from a GGA calculation without the Coulomb $U$, and that from a GGA+$U$ calculation ($U=3$ eV and $J=0.6$ eV are used for any GGA+$U$ calculation mentioned henceforth). In both pictures, the Pt manifold, predominantly $5d$ in character in the region shown, ends near the $-2$ eV mark and is fully occupied; such is the general case for $4d$ and $5d$ metals in the 115s and 218s. Thus Pt is neutral or possibly slightly negatively charged in these compounds. In the GGA case, the two large Pu peaks correspond to the 5$f_{5/2}$, 5$f_{7/2}$ SO splitting of very narrow $f$ bands. The peaks are separated by roughly 1 eV, which is the expected splitting level for Pu compounds.

With the addition of $U$, the Pu peaks each split into multiple smaller peaks. The occupied peak broadens to span a range of 1.5 eV; the unoccupied peak shifts 0.8 eV to the right and creates a trail of $f$~character up to above 4 eV. The Pu bands widen as a result of the on-site Coulomb repulsion and exchange interaction $J$. The DOS at $\varepsilon_F$ is $N(0) = 6.32$~eV$^{-1}$ (down from the GGA DOS of 9.07~eV$^{-1}$), which gives a noninteracting electronic specific heat coefficient of 15~mJ/mol-K$^2$. Comparison with the experimentally measured Sommerfeld coefficient of 250~mJ/mol K$^2$ gives a mass renormalization of $\sim$17, which cannot be captured by our static mean-field calculations. Dynamical correlations as in the Kondo effect are responsible for this discrepancy, as observed for the other Pu compounds in this family.

\begin{figure}[tbh]
\begin{center}
\includegraphics[draft=false,width=\columnwidth,clip]{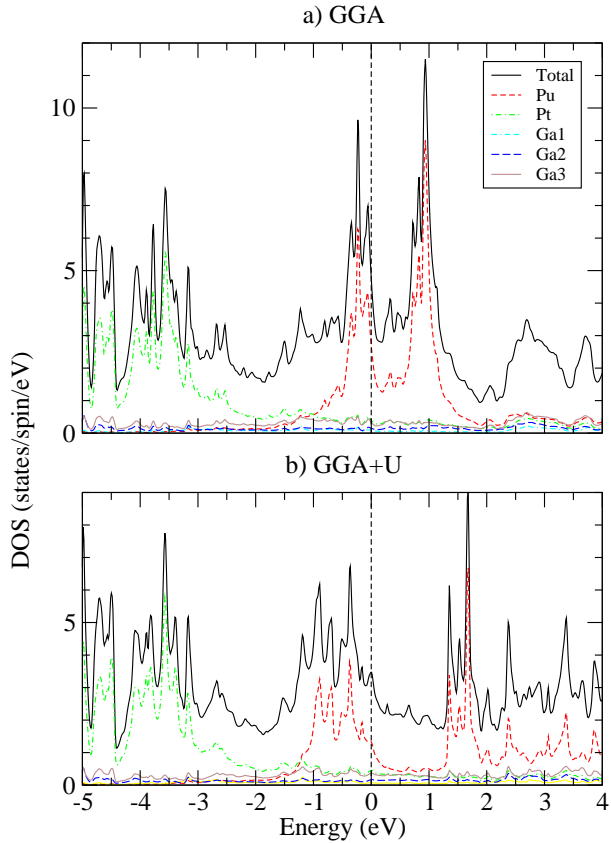}
\end{center}
\caption{(Color online) Total and partial DOSs of PM PuPt$_2$Ga$_7$ from a) GGA and b) GGA+$U$ ($U=3$ eV, $J=0.6$ eV) calculations.}
\label{ga-dos}
\end{figure}

Fig.~\ref{ga-dos} provides the DOS of nonmagnetic PuPt$_2$Ga$_7$ from GGA and GGA+$U$ calculations. As in PuPt$_2$In$_7$, the Pt $5d$ states are filled and the Pu $5f$ peaks, which are located between $-1$ and +1.5 eV before the implementation of $U$, spread to a wider range when $U$ is turned on. The bands are generally broader compared to PuPt$_2$In$_7$, due to the lattice constants of PuPt$_2$Ga$_7$ (the smaller volume overrides the shortness of the Ga wavefunction). When the states near $\varepsilon_F$ are decomposed into their total angular momentum quantum numbers $m_j$, we find that the Pu states with $m_j = \pm 3/2$ dominate the Fermi energy. This is consistent with the idea that the most relevant hybridization will be between Pu and its nearest neighbors, which are not the in-plane but rather out-of-plane In atoms.

\begin{figure}[tbh]
\begin{center}
\includegraphics[draft=false,width=\columnwidth,clip]{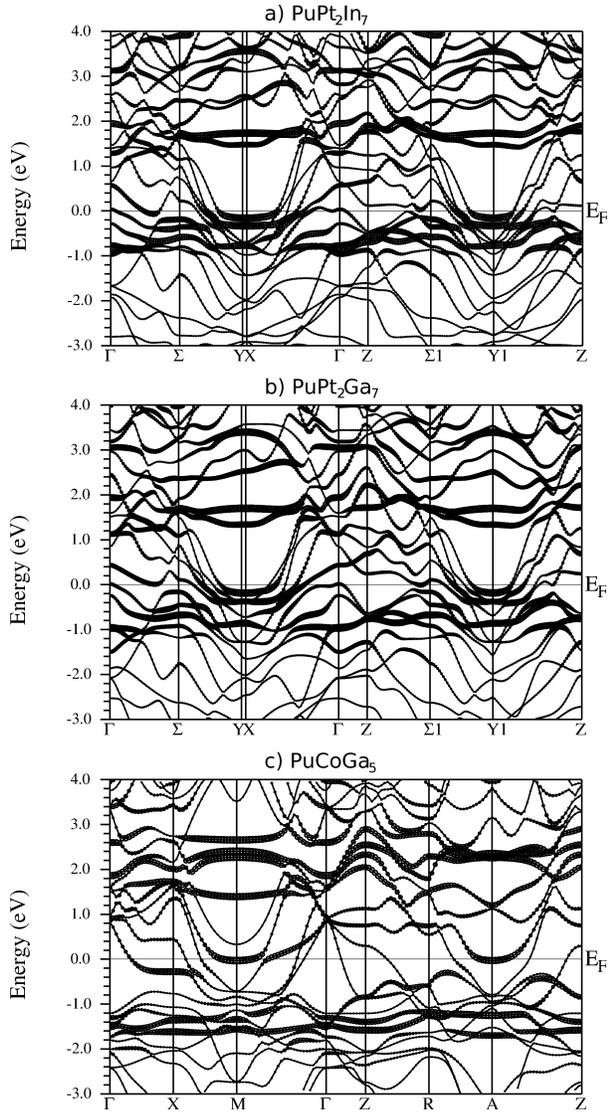}
\end{center}
\caption{GGA+$U$ band structures of PM (a) PuPt$_2$In$_7$, (b) PuPt$_2$Ga$_7$, and (c) PuCoGa$_5$, with $f$-weight fatbands.}
\label{bs}
\end{figure}

\begin{figure}[tbh]
\begin{center}
\includegraphics[draft=false,width=\columnwidth]{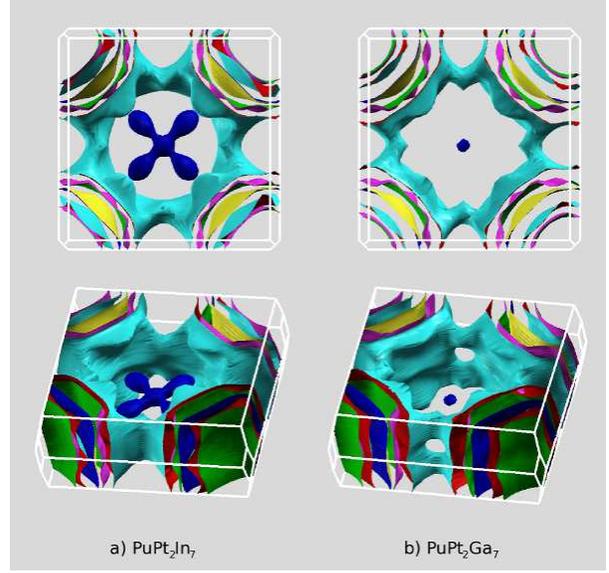}
\end{center}
\caption{(Color online) Calculated FSs of (a) PuPt$_2$In$_7$ and (b) PuPt$_2$Ga$_7$ in the GGA+$U$ scheme. $\Gamma$ is located in the center of the unit cell. For clarity, the 3D FSs are reproduced in the bottom figures.}
\label{fs}
\end{figure}

\begin{figure}[tbh]
\begin{center}
\includegraphics[draft=false,width=\columnwidth]{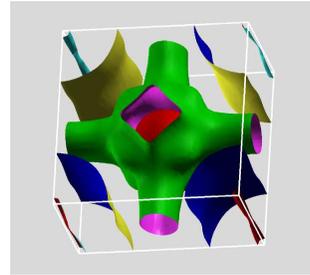}
\end{center}
\caption{(Color online) Calculated FSs of PuCoGa$_5$ in the GGA+$U$ scheme. $\Gamma$ is located in the center of the unit cell.}
\label{pucoga5-fs}
\end{figure}

The band structures of PM PuPt$_2$In$_7$, PuPt$_2$Ga$_7$, and PuCoGa$_5$ obtained from GGA+$U$ calculations are shown in Fig.~\ref{bs}. The thickness of a band corresponds to the weight of the $f$ orbital. In PuPt$_2$In$_7$ and PuPt$_2$Ga$_7$, the $f_{5/2}$ and $f_{7/2}$ fatbands are visible right below and 1.5 eV above $\varepsilon_F$, respectively. In PuCoGa$_5$, most of the $f_{5/2}$ states are shifted downward, but the rest are concentrated at the Fermi level in relatively dispersionless form. The highly dispersive band, which spans almost 2 eV from Z to $\Gamma$ and crosses the Fermi energy in PuCoGa$_5$, barely reaches $\varepsilon_F$ in the 127s and creates a small hole Fermi surface pocket at the center of the zone (see Fig.~\ref{fs}). This indicates a reduction in dimensionality when going from the 115 to the 127, but the reduction effect is not as obvious when looking at the FSs as a whole (compare Figs.~\ref{fs} and \ref{pucoga5-fs}). We therefore used {\sc WIEN2k} to calculate the plasma frequency ratio $\omega_{p,xx}/\omega_{p,zz}$ ($ = \langle v_x^2 \rangle^{1/2}/\langle v_z^2 \rangle^{1/2}$) of PuPt$_2$In$_7$, PuPt$_2$Ga$_7$, PuCoIn$_5$, and PuCoGa$_5$, and they are 2.34, 3.22, 1.46, and 1.68, respectively. As expected, all four ratios are $> 1$. The larger value of PuPt$_2$Ga$_7$ (PuCoGa$_5$) indicates two-dimensionality is enhanced when compared to PuPt$_2$In$_7$ (PuCoIn$_5$), despite its smaller volume. This indeed demonstrates that the 127 compounds are electronically more anisotropic than the 115 compounds. In addition, the Ga compounds, despite their smaller structure, are slightly more 2D than their In analogs.

If, as in the case of the Ce-based superconductors, the presence of superconductivity relies on the proximity to an antiferromagnetic state, we would like to know the relative degree of localization in the various Pu-115, -127, and -218 compounds. From the DFT calculations, we can get an estimate for the relative strength of the $c$-$f$ hybridization. We take the $f$-electron density within the Pu muffin-tin sphere to be inversely related to the strength of hybridization. For identically sized MT spheres (3.1), we find $f$-occupations of 5.24 for both PuCoIn$_5$ and PuPt$_2$In$_7$ and 5.14 for both PuCoGa$_5$ and PuPt$_2$Ga$_7$. Thus, we obtain that the In compounds are less hybridized than the Ga analogs. This result alone does not indicate the degree of localization. However, dynamical mean-field theory (DMFT) calculations show that the more weakly hybridized PuCoIn$_5$ indeed results in a smaller Kondo scale, $T_0$, relative to PuCoGa$_5$, and hence can be considered as more localized.\cite{Zhu} Thus, we can now equate the relative degree of hybridization with the relative degree of localization, and we conclude that PuCoIn$_5$ and PuPt$_2$In$_7$ have a similar degree of localization which is stronger than the more itinerant PuCoGa$_5$ and hypothetical PuPt$_2$Ga$_7$. As a result, since PuCoIn$_5$ is non-magnetic it is not surprising that PuPt$_2$In$_7$ is also non-magnetic.

\begin{figure}[tbh]
\begin{center}
\includegraphics[draft=false,width=\columnwidth,clip]{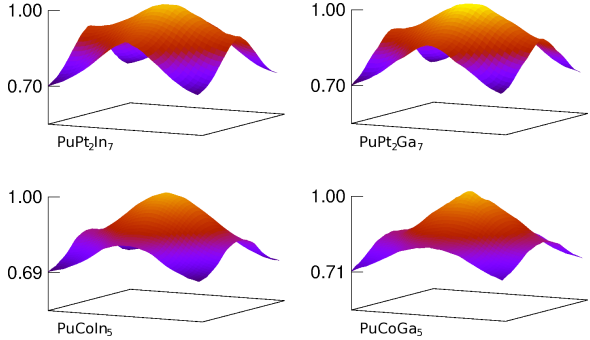}
\end{center}
\caption{(Color online) $f$-weighted normalized noninteracting spin susceptibilities $\tilde\chi$ of Pu-based compounds along the $q_xq_y$ plane in the conventional Brillouin zone for $q_z = 0.5$. $\mathbf q=(0,0,0)$ are at the corners.}
\label{suscs}
\end{figure}

The role of the electronic structure in determining superconductivity depends on the mechanism. It has been argued that in some cases, superconductivity can be driven by Fermi surface nesting. Nesting, which indicates instability in the FS, can give rise to a spin density wave or charge density wave. In a BCS-like mechanism, even if the pairing fluctuations do not originate directly from a FS instability, the electronic structure will at a minimum determine the superconducting gap symmetry, as well as the character of the interaction. In the Pu-218s, Elgazaar {\it et al.} have argued that the additional FS sheets may provide sufficient differences to suppress the occurrence of superconductivity.\cite{elgazzar} To see if there is any nesting present in the Pu-127s, we have used the GGA band structures to calculate the real part of the constant-matrix-element noninteracting susceptibility for PuPt$_2$In$_7$ and PuPt$_2$Ga$_7$ (see Fig.~\ref{suscs}). In the interest of finding nesting features that are unique to the superconductors, we calculated the susceptibilities of PuCoGa$_5$ and PuCoIn$_5$ as well. The generalized susceptibility is
\begin{equation*}
\chi(\mathbf q) = -\sum_{\alpha\beta\mathbf k} \frac{f(\varepsilon_{\alpha,\mathbf k}) - f(\varepsilon_{\beta,\mathbf{k+q}})}{\varepsilon_{\alpha,\mathbf k} - \varepsilon_{\beta,\mathbf{k+q}} + i\delta},
\end{equation*}
where $f$ denotes the Fermi distribution function, $\varepsilon_{\alpha,\mathbf k}$ is the energy dispersion, and $\alpha$ and $\beta$ are band indices. Alongside the conventional $\chi(\mathbf q)$, we also calculated the susceptibility incorporating the relative weight of the Pu $f$~orbital, so as to pick out the attributes dominated by Pu $f$~character. In the style of Mazin as in Ref.~\onlinecite{mazin}, the \textit{weighted} susceptibility $\tilde\chi(\mathbf q)$ is
\begin{equation*}
\tilde\chi(\mathbf q) = -\sum_{\alpha\beta\mathbf k} \frac{f(\varepsilon_{\alpha,\mathbf k}) - f(\varepsilon_{\beta,\mathbf{k+q}})}{\varepsilon_{\alpha,\mathbf k} - \varepsilon_{\beta,\mathbf{k+q}} + i\delta} W_{\alpha,\mathbf k}W_{\beta,\mathbf{k+q}},
\end{equation*}
where $W$ is the weight of the $f$~orbital. Shown in Fig.~\ref{suscs} are the $f$-weighted $\tilde\chi(\mathbf q)$, which are normalized and plotted along the $q_xq_y$ plane for $q_z = 0.5$, of the four compounds. In each case, the non-weighted $\chi(\mathbf q)$ looks almost identical to its weighted counterpart, demonstrating that the weights of other atoms and orbitals were negligible to begin with. PuPt$_2$In$_7$ and PuPt$_2$Ga$_7$ have similar-looking susceptibility plots, as do PuCoGa$_5$ and PuCoIn$_5$. Moreover, the susceptibilities of the 115s are not very dissimilar to those of the 127s. The primary difference is that the 127s feature elevated values along $(0.5,q_x)$ [and equivalently, $(q_y,0.5)$], which can also be seen, to a much lesser degree, in PuCoIn$_5$. The peak-like character is most pronounced for PuCoGa$_5$, which has the highest $T_c$ of the four compounds.

\begin{figure}[tbh]
\begin{center}
\includegraphics[draft=false,width=\columnwidth,clip]{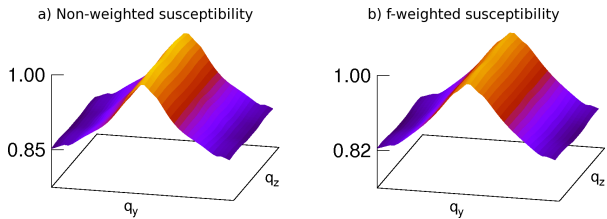}
\end{center}
\caption{(Color online) Normalized a) $\chi$ and b) $\tilde\chi$ of PuCoGa$_5$ in the conventional Brillouin zone for $q_x = 0.5$. $\mathbf q=(0,0,0)$ are at the corners. Susceptibilities of the other three Pu compounds look qualitatively equivalent.}
\label{mtnridge}
\end{figure}

Wang et al.\cite{wang} noted two peaks in PuCoGa$_5$'s $\chi(\mathbf q)$, at $\mathbf q = (0.5,0.5,0)$ and $\mathbf q = (0.5,0.5,0.5)$. More accurately, the two peaks are part of a relatively broad ridge that, when plotted on the $q_yq_z$ (or, equivalently, $q_xq_z$) plane, spans all the way in the $q_z$ direction. This ridge is seen in all four Pu compounds, and is plotted in Fig.~\ref{mtnridge} for PuCoGa$_5$. When $\chi$ and $\tilde\chi$ are plotted along a $q_x q_y$ plane for any $q_z$, the apex appears at the corner of the Brillouin zone ($q_x = q_y = 0.5$), as can be seen in Fig.~\ref{suscs}. That there is little variation in the landscape when varying $q_z$ indicates a truly 2D topography in the susceptibility for both Pu-115s and Pu-127s.

When it comes to the source of the maxima ($0.5,0.5,q_z$), in all cases, the biggest contribution is interband nesting involving the largest FS sheet. In the 115s, the large sheet connects with the larger of the two 2D cylinders (Fig.~\ref{pucoga5-fs}); in the 127s, it maps onto the two largest cylinders (Fig.~\ref{pucoga5-fs}), where nesting with the bigger of the two cylinders is stronger than nesting with the smaller, by 7\%/20\% for PuPt$_2$In$_7$/PuPt$_2$Ga$_7$. Nesting between the large sheet and the cylinder accounts for, on average, 32\% of the susceptibility strength of PuCoGa$_5$, while that factor is only 24\% for PuCoIn$_5$. Nestings between the large sheet and the two larger cylinders collectively account for 28\% for PuPt$_2$Ga$_7$ and 26\% for PuPt$_2$In$_7$. PuPt$_2$In$_7$ and PuPt$_2$Ga$_7$ have similar susceptibility plots, as do PuCoGa$_5$ and PuCoIn$_5$, and even the individual band-decomposed $\chi_{\alpha\beta}$'s are consistent throughout the compounds. This demonstrates that the type of ligand atom has very little influence on the shape of $\chi(\mathbf q)$.

What do these calculations tell us about superconductivity? The virtually identical Fermi surfaces, $\chi(\mathbf q)$ plots, and $\tilde\chi(\mathbf q)$ plots at the DFT level for PuCoIn$_5$ and PuCoGa$_5$ whose superconducting $T_c$ differs by nearly an order of magnitude, suggests that an additional energy scale must be important. The most likely candidate is the Kondo energy scale, $T_0$, extracted from either specific heat measurements or DMFT calculations. As mentioned above, earlier DMFT work on these two compounds shows that the hybridization strength inferred from DFT calculations can predict the relative trend of $T_0$ between various Pu-based family members.\cite{Zhu} Consequently, our work shows that $T_0$ is similar for PuCoIn$_5$ and PuPt$_2$In$_7$ as well as between PuCoGa$_5$ and PuPt$_2$Ga$_7$. Thus, we naively expect the scale of $T_c$ for PuPt$_2$In$_7$ to be similar to PuCoIn$_5$. As a result, it is surprising that PuPt$_2$In$_7$ is not superconducting, especially given the similarity of the susceptibility between the Pu-115's and the Pu-127's. Of course, subtle differences do exist in $\chi(\mathbf q)$ which may be sufficient to drive $T_c$ below 2 K in PuPt$_2$In$_7$.

\section{Conclusion}
We have reported the properties of PuPt$_2$In$_7$ a structurally more 2D version of the known Pu-based superconductors. The gross similarities in structure and FSs between PuPt$_2$In$_7$ and the other known Pu-based superconductors suggest that PuPt$_2$In$_7$ may be a likely candidate to find superconductivity. While neither superconductivity nor magnetic order was observed down to 2 K, our calculations suggest possible ordering below 2 K. Our study of a hypothetical PuPt$_2$Ga$_7$ reveals strong similarites to PuPt$_2$In$_7$ and PuCoGa$_5$, suggesting that it is a promising candidate to find superconductivity if it can be synthesized. More work is needed to explore these various possibilities.

\section{Acknowledgments}
This work was supported by DOE grant DE-FG02-04ER46111,
the Strategic Sciences Academic Alliance Program under grant
DE-FG03-03NA00071, and by DOE SciDAC Grant No. DE-FC02-06ER25794.  Work at Los Alamos was performed under the auspices of the U.S. DOE, Office of Science, Division of Materials Sciences and Engineering, and supported in part by the Laboratory Directed Research and Development program.  Work at Lawrence Berkeley National Laboratory was supported by the U.S.
Department of Energy (DOE), Office of Basic Energy Sciences (BES) under
Contract No.  DE-AC02-05CH11231.  X-ray absorption data were collected at SSRL,
a national user facility operated by Stanford University on behalf of the
DOE/BES.

\end{document}